\documentstyle[twocolumn,aps]{revtex}
\input psfig

\begin{document}
\draft

\twocolumn[\hsize\textwidth\columnwidth\hsize\csname @twocolumnfalse\endcsname

\title{Spectral Function and Self-Energy of the One-Dimensional 
Hubbard Model in the $U\rightarrow\infty$ Limit}
\author{Fr\'ed\'eric Mila$^{a}$ and 
 Karlo Penc$^{b}$\cite{*} 
  }

\address{
     $(a)$ Laboratoire de Physique Quantique, Universit\'e Paul Sabatier,
     31062 Toulouse (France) \\
     $(b)$ Max-Planck-Institut f\"ur Physik komplexer Systeme, 
          Bayreuther Str.~40, 01187 Dresden (Germany) \\
}

\maketitle

\begin{abstract}
The interpretation of the $k$ dependent spectral functions of the  
one--dimensional, infinite $U$ Hubbard model obtained by using the 
factorized wave--function of Ogata and Shiba is revisited.
The well defined feature which appears in addition to low energy features 
typical of Luttinger liquids, and which, close to the Fermi
energy, can be interpreted as the shadow
band resulting from $2k_F$ spin fluctuations, is further investigated.
A calculation of the self-energy
shows that, not too close to the Fermi energy, this feature corresponds to 
a band, i.e. to a solution of the Dyson equation 
$\omega - \epsilon(k) - \rm{Re} \Sigma (k,\omega) =0$.
\end{abstract}


\vskip2pc]

\narrowtext
In the study of systems of interacting electrons, the dynamical correlation
functions play a central role because they correspond to the response functions
that can be measured in several experiments (infrared reflectivity, inelastic
neutron scattering, nuclear magnetic resonance, etc...) The most basic of these
correlation functions is the time-ordered Green's function, 
which can be defined
by its spectral representation according to 
\begin{equation}
  G(k,\omega)  = 
  \int_{\mu}^{+\infty} {\rm d} \omega' 
  \frac{A(k,\omega')}{\omega \!-\! \omega' \!+\! i\delta}
 +  \int_{-\infty}^{\mu} {\rm d} \omega' 
  \frac{B(k,\omega')}{\omega \!-\! \omega' \!-\! i\delta} 
 \label{eq:GAB}
\end{equation}
The spectral functions $ A(k,\omega)$ and $B(k,\omega)$ are defined by 
\begin{eqnarray}
  A(k,\omega) &=&
 \sum_{f,\sigma} 
  \left| \langle f,N \!+\! 1| c^\dagger_{k,\sigma} |0,N\rangle \right|^2
  \delta(\omega \!-\! E^{N+1}_f \!+\! E^N_0) 
  \>\nonumber\\
  B(k,\omega) &=&
 \sum_{f,\sigma} 
  \left| \langle f,N \!-\! 1| c^{\phantom{\dagger}}_{k,\sigma} 
  |0,N\rangle \right|^2
    \delta(\omega \!-\! E^{N}_0 \!+\! E^{N-1}_f) 
  \>
\nonumber
\end{eqnarray}
in the standard notation,
and in principle they can 
be measured in angular-resolved inverse photoemission and 
photoemission experiments, respectively. 
The effect of electron-electron interactions on the spectral functions of
the three-dimensional Coulomb gas has been
investigated in much detail several years
ago\cite{hedin}. With respect to the simple $\delta(\omega - \varepsilon_k)$ 
structure of the
non-interacting case, the main differences are: i) A shift of the energy of the 
quasi-particle band; ii) A broadening of the quasi-particle peak; iii) A 
renormalization of the weight of the
quasi-particle peak, compensated by the appearance of an incoherent background;
iv) The presence of another band - a plasmon band - due to 
the long-range nature of
the Coulomb potential. For a short-range repulsion, such as 
the on-site interaction
of the Hubbard model, there is no plasmon band, and the spectral function is
expected to have only one well-defined feature, the quasi-particle band, on top
of an incoherent background. The validity  of this simple 
picture in lower dimensional systems 
is currently under intense discussion. In two dimensions(2D), 
there are at least two important issues:
1. the actual existence of a quasiparticle peak, and 2. the generation by
antiferromagnetic short-range fluctuations close to half-filling
of additional well-defined features 
in the spectral functions, 
called generically ``shadow bands''\cite{kampf}.
The absence 
of exact results in 2D for correlated models makes the
interpretation of the experimental results quite difficult. For high--T$_c$
cuprates, these issues are  still controversial\cite{2D}.

In 1D, a number of analytical approaches are available. For instance,
models having both charge and spin low-lying excitations belong to the
universality class of the Luttinger liquid, and 
bosonization gives an accurate description of their low-energy properties.
The consequence for the spectral functions is that there is no
quasi-particle peak but two divergences due to spin-charge
separation\cite{meden}. 
A lot of insight has also been obtained by the Bethe ansatz solution of
integrable models, the prototype being
the Hubbard model defined by:
\begin{equation}
  {\cal H}= -t \sum_{i,\sigma} 
   \left(
    c^\dagger_{i,\sigma} c^{\phantom{\dagger}}_{i+1,\sigma}
   + h.c. \right)
   + U \sum_{i} n_{i,\uparrow} n_{i,\downarrow}
\end{equation}
where $c^\dagger_{i,\sigma}$ and $c^{\phantom{\dagger}}_{i,\sigma}$ are
electron creation and annihilation operators, and $n_{i,\sigma}=
c^\dagger_{i,\sigma} c^{\phantom{\dagger}}_{i,\sigma}$ is the density operator.
The Bethe ansatz wave-functions are so complicated that it is not possible to
use them to calculate correlation functions. However, Ogata and Shiba showed a
few years ago that, in the limit $U\rightarrow +\infty$, the ground-state 
wave function
can be written as a product of a spinless fermion wave--function 
$|\psi\rangle$ and a squeezed spin wave--function $|\chi\rangle$, and that this
wave function allows a very precise evaluation of the momentum distribution
function\cite{shiba}. More recently, Penc et al\cite{penc1,penc2} 
used the same representation 
for the excited
states to calculate the spectral functions $A(k,\omega)$ and $B(k,\omega)$. The
results for $n=1/4$ are shown in Fig. 1.

There are several interesting features 
to notice. In the low energy region near $k_F$ we can identify three
structures. For $k<k_F$ there are maxima
at $\omega=u_c (k-k_F)$ and $\omega\simeq 0$ 
and a lot of spectral weight 
between them. 
There is also a small weight
appearing on the other side of the Fermi energy for $\omega>-u_c(k-k_F)$. 
If we remember
that the spin velocity $u_s$ vanishes for the infinite $U$ Hubbard model, all
these features are qualitatively consistent with the Luttinger liquid 
calculations\cite{meden}. 
The dispersion of the charge part is exactly given
by $E(k)=-2t \cos(|k|+k_F)$. 

\begin{figure}[hp]
\centerline{\psfig{figure=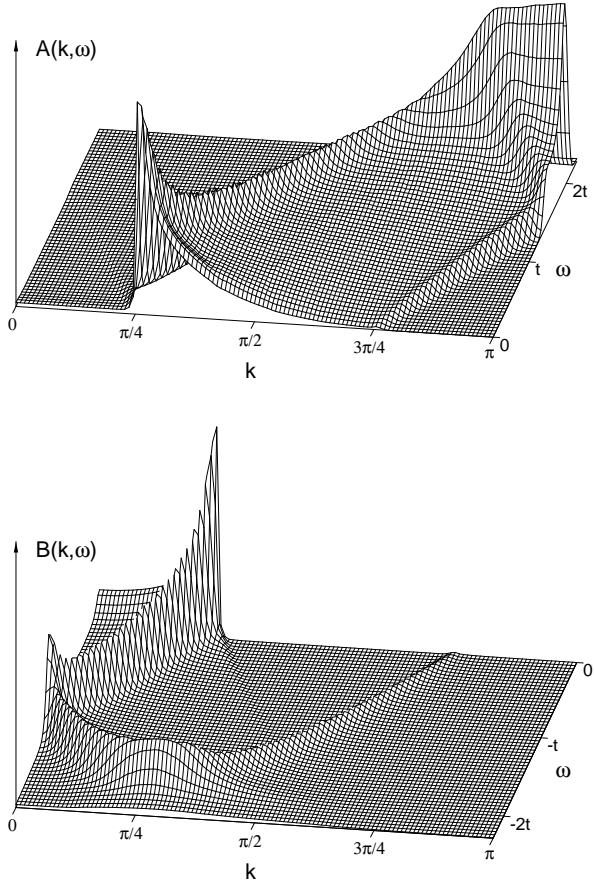,width=8.0cm,angle=0}}
\vspace{0.5cm}
\caption{
Spectral functions of the $U\rightarrow +\infty$ Hubbard model
for $L=228$ sites and $N=114$ electrons ($n=1/4$) with Fermi momentum
$k_F=\pi/4$.}
\label{fig:3d}
\end{figure}

However, the Luttinger liquid picture does not exhaust the features of the
spectral function of Fig.~1. For larger energies, or away from
$k_F$,
there is a well defined band--like structure  with
considerable spectral weight and a dispersion given by 
$E(k)=-2t \cos(-|k|+k_F)$.
This band crosses the Fermi energy at $3k_F$ and has been interpreted as a
shadow-band originating from the diverging $2k_F$ spin fluctuations. 
The presence of a shadow-band in a 1D lattice model has been first reported on
the basis of exact diagonalizations of finite clusters by Haas and
Dagotto\cite{haas}. More recently, a systematic study of various extensions of
the infinite $U$ Hubbard model to finite repulsions has shown that these
features are robust down to reasonable values of the parameters\cite{favand}.

While there
is little doubt that this physical picture is essentially correct, the
interpretation should be made more precise in two respects.
First of all, the concept of a shadow band is no longer well defined 
below the energy where the charge part of the main band crosses $k=0$. In fact,
if we think in terms of the whole Brillouin zone ($-\pi < k < \pi$), what we
call a shadow-band for $k>0$ is continuously connected to the main band for
$k<0$. So what we have called a shadow band is probably better thought of as a
band {\it per se} which, close to the Fermi level, can be interpreted as a 
shadow band.

Second, the term band has a precise meaning in the context of perturbation
theory. A feature of the spectral function is called a band if it corresponds to
a solution of the Dyson equation 
$\omega - \epsilon(k) - \rm{Re} \Sigma (k,\omega) =0$, where 
$\epsilon(k)=-2t\cos k$ is the non-interacting dispersion while
the self-energy
$\Sigma (k,\omega)$ is implicitly defined by 
$G(k,\omega)=1/(\omega-\epsilon(k)-\Sigma (k,\omega))$ and can be obtained from
$G(k,\omega)$ as $\Sigma (k,\omega)=\omega-\epsilon(k)-G^{-1}(k,\omega)$.
For 1D systems, the perturbation expansion diverges notoriously, and, even for
small $U/t$, $\Sigma (k,\omega)$ cannot be calculated perturbatively. However,
having obtained $A(k,\omega)$ and $B(k,\omega)$ in an essentially exact way, we
can use Eq. (1) to calculate the time ordered Green's function, and from it the
self-energy. The results for the real part and the imaginary part of the
self-energy for $k=\pi/3$ and $k=2\pi/3$ are given in Fig. 2.

To discuss the relationship between the spectral functions and the self-energy,
it is useful to write $A(k,\omega)$ and $B(k,\omega)$ in terms of 
$\Sigma (k,\omega)$ as
\begin{eqnarray}
  A(k,\omega) &=&
\pi^{-1}{-\rm{Im}\Sigma (k,\omega) \over (\omega-\epsilon(k)-\rm {Re} 
\Sigma (k,\omega))^2 +
(\rm{Im}\Sigma (k,\omega))^2}
\nonumber\\
  B(k,\omega) &=&
\pi^{-1}{\rm{Im}\Sigma (k,\omega) \over (\omega-\epsilon(k)-\rm  {Re}
\Sigma (k,\omega))^2 +
(\rm{Im}\Sigma (k,\omega))^2}
\nonumber
\end{eqnarray}

Let us start with $k=2\pi/3$. As for a non-interacting system, the Dyson
equation has only one solution corresponding to the intersection of the straight
line with the real part of $\Sigma (k,\omega)$. This corresponds to the charge
part of the main excitation band. The spin part, which is pinned at the Fermi
energy in the present case, corresponds to a vertical slope at $\omega=0^+$ 
in the imaginary part of the self-energy, as well as in the spectral function 
$A(k,\omega)$. The ``shadow'' band, which appears at negative energies for that
wave-vector, does {\it not} correspond to a solution of the Dyson equation, but
to a maximum of the imaginary part that occurs in a region where
$\omega-\epsilon(k)-\rm {Re} \Sigma(k,\omega)$ is large, so that the 
denominator of 
$A(k,\omega)$ is dominated by $(\omega-\epsilon(k)-\rm {Re}\Sigma(k,\omega))^2$.

For $k=2\pi/3$, there are three solutions to the Dyson equation. As in the
previous case, the solution for positive $\omega$ corresponds to the charge part
of the main band, and the spin part also gives rise to a singularity at 
$\omega=0^+$. For negative frequencies, the situation is very similar to that of
the plasmon band of the 3D Coulomb gas\cite{hedin}. 
Decreasing $\omega$ starting from the
Fermi energy, the first solution of the Dyson equation coincides with a maximum
of $\rm {Im} \Sigma(k,\omega)$ and does not give rise to any feature in the
spectral function. On the contrary, the other solution occurs in a region where 
$\rm {Im} \Sigma(k,\omega)$ is not particularly big. This solution gives rise to
a feature in the spectral function that corresponds to what we have called 
the ``shadow'' band.
 
\vskip.5cm
\begin{figure}[hp]
\centerline{\psfig{figure=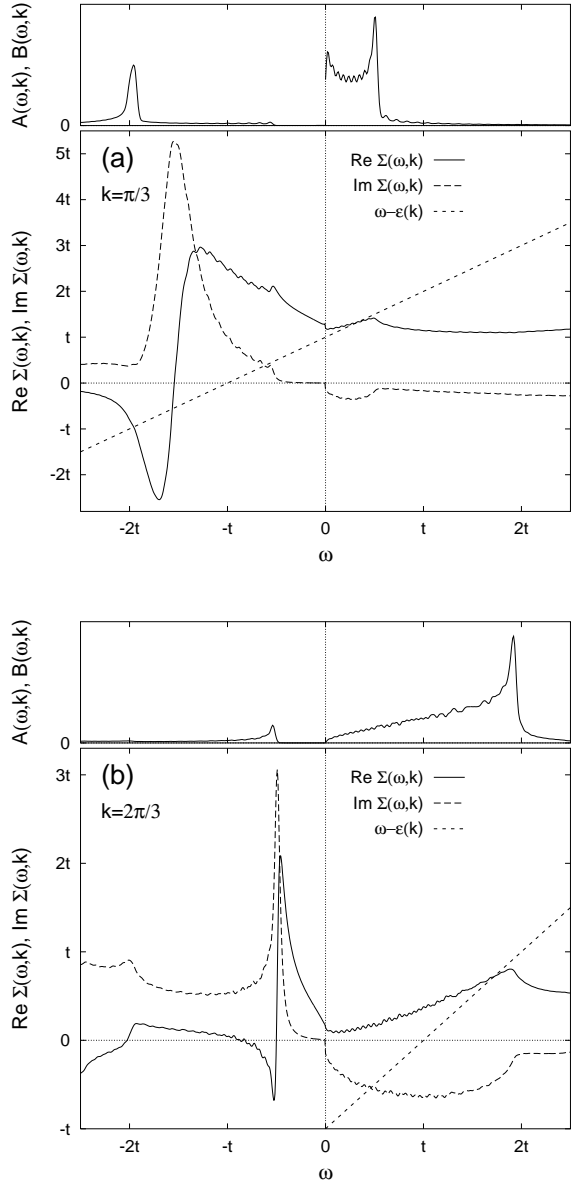,width=7.5cm,angle=0}}
\vspace{.5cm}
\caption{Frequency dependence of the spectral functions and of the
real and imaginary parts of the self-energy of the quarter-filled, 
$U\rightarrow +\infty$ Hubbard model for selected values of the momentum. 
The straight line corresponds to
$\omega-\epsilon(k)$. a) $k=\pi/3$; b) $k=2\pi/3$. }
\label{fig:2d}
\end{figure}

Coming back to the second point raised above, we see that the feature that
crosses the Fermi energy at $3k_F$ corresponds to a band, i.e. to a solution of
the Dyson equation, only far enough from the Fermi surface. In that case however
it can't really be interpreted as a ``shadow'' band because it cannot be
connected to the main band by a vector $2k_F$. Close enough to the Fermi
energy, this feature can really be thought of as a shadow of the main band
because it really follows the charge part of the main band at a distance $2k_F$
apart. However, this feature does not correspond any more to a solution of the
Dyson equation, and strictly speaking, it should not be called a band. 

To summarize, this work is the first attempt at using the self-energy 
to interpret the spectral functions of the Hubbard model that have been 
obtained in an essentially exact way in the $U\rightarrow +\infty$ limit. 
Extracting the self-energy from the spectral functions numerically has already
allowed us to draw an interesting conclusion concerning the origin of the
feature that was previously interpreted as a shadow band, namely that this
feature only corresponds to a solution of the Dyson equation far enough from the
Fermi level. The next step, 
a detailed analysis of the behaviour of the self-energy close to
($\omega=0,k=k_F,3k_F$), is in progress. It might help clarify the 
still controversial problem
of the spectral function in 2D in the presence of almost perfect nesting.

We thank H. Shiba and A.-M. Tremblay
for very useful discussions.


\begin{references}

\bibitem[*]{*} 
  On leave from Research Institute for Solid State Physics, Budapest, Hungary.
  
\bibitem{hedin} L. Hedin and S. Lundqvist, Solid State Physics {\bf 23}, 1
(Academic, New York, 1969).

\bibitem{kampf} A. P. Kampf and J. R. Schrieffer, Phys. Rev. B {\bf 42}, 7967
(1990).

\bibitem{2D} For recent developments in 2D, see S. Haas, A. Moreo and
E. Dagotto, Phys. Rev. Lett. {\bf
74}, 310 (1995); R. Preuss, W. Hanke, W. von der Linden, Phys. Rev. Lett. 
{\bf 75}, 1344 (1995); A.
Chubukov, Phys. Rev. B {\bf 52}, R3840 (1995); M. Langer, J. Schmalian, S.
Grabowski, K. H. Bennemann, Phys. Rev. Lett. {\bf 75}, 4508 (1995); J.
Schmalian, M. Langer, S. Grabowski, K. H. Bennemann, Phys. Rev. B {\bf 54}, 4336
(1996); D. Duffy and A. Moreo, Phys. Rev. B {\bf 52}, 15607 (1995).
  
\bibitem{meden}
  V. Meden and K. Sch\"onhammer, Phys. Rev. B {\bf 46}, 15753 (1992);
  K. Sch\"onhammer and V. Meden, ibid. {\bf 47}, 16205 (1993); 
  J. Voit, ibid.  {\bf 47}, 6740 (1993).

\bibitem{shiba}
  M. Ogata and  H. Shiba,  Phys. Rev. B {\bf 41}, 2326 (1990);
  M. Ogata, T. Sugiyama and H. Shiba, ibid. {\bf 43}, 8401 (1991); 
  
\bibitem{penc1} K. Penc, F. Mila, H. Shiba, Phys. Rev. Lett. {\bf
75}, 894 (1995).

\bibitem{penc2} K. Penc, K. Hallberg, F. Mila, H. Shiba, Phys. Rev. Lett. {\bf
77}, 1390 (1996).

\bibitem{haas} S. Haas and E. Dagotto, Phys. Rev. B {\bf 52}, R14396 (1995).

\bibitem{favand} J. Favand, S. Haas, K. Penc, F. Mila and E. Dagotto, Phys. Rev.
B, Rapid Comm., in press.

\end{references}
\end{document}